\documentclass {aa}  

\usepackage{graphicx}
\usepackage{tikz}
\usepackage{bm}
\usepackage{txfonts}
\begin{document} 

   \title{Investigation of the bi-drifting subpulses of radio pulsar B1839--04 utilising the open-source data-analysis project PSRSALSA}

   \titlerunning{Investigation of the bi-drifting subpulses of PSR B1839--04 utilising PSRSALSA}

   \author{P. Weltevrede
          \inst{1}
          }

   \institute{Jodrell Bank Centre for Astrophysics, The University of Manchester, Alan Turing Building, Manchester, M13 9PL, United Kingdom\\
              \email{Patrick.Weltevrede@manchester.ac.uk}
             }

   \date{Received December 13, 2015; accepted March 2, 2016}

 
  \abstract
   {}
   {The usefulness and versatility of the PSRSALSA open-source pulsar 
data-analysis project  is
demonstrated through an analysis of the radio pulsar B1839--04.
This study focuses on the phenomenon of bi-drifting, an effect where the drift
direction of subpulses is systematically different in different pulse profile
components. Bi-drifting is extremely rare in the pulsar population, and the theoretical implications are discussed
after comparing B1839--04 with the only other known bi-drifter.}
   {Various tools in PSRSALSA, including those allowing quantification of periodicities in the subpulse modulation, their flux distribution, and polarization properties, are exploited to obtain a comprehensive picture of the radio properties of PSR B1839--04. In particular, the second harmonic in the fluctuation
     spectra of the subpulse modulation is exploited to convincingly demonstrate the existence of
     bi-drifting in B1839--04. Bi-drifting is confirmed with a completely independent method
     allowing the average modulation cycle to be determined. Polarization measurements were used
     to obtain a robust constraint on the magnetic inclination angle.
   }
   {The angle between the rotation and magnetic axis is found to be smaller than $35\degr$. Two distinct
     emission modes are discovered to be operating, with periodic subpulse modulation
being present only during the weaker mode. Despite the variability of the
modulation cycle and interruption by mode-changes, the modulation pattern responsible for the
bi-drifting is strictly phase locked over a timescale of years such that the variability is identical in the different components.

   }
   {The phase locking implies that a single physical origin is responsible for
both drift directions. Phase locking is hard to explain for many models, including those specifically proposed in the
literature to explain bi-drifting, and they are therefore shown to be implausible. It is argued that within the framework of circulating beamlets, bi-drifting could occur
if the circulation were severely distorted, possibly by distortions in the magnetic field.
}


   \maketitle
%

\section{Introduction}

The aim of the discussion in this manuscript is two-fold. First of
all, the unusual radio behaviour of \object{PSR B1839$-$04}, especially at the
individual pulse level, is presented, and the consequences for our
theoretical understanding of the mechanism responsible for the
emission are discussed. Secondly, the software tools that were used to
obtain these results, which are part of a new open-source
data-analysis project called PSRSALSA\footnote{The latest version of PSRSALSA and a tutorial can be downloaded from https://github.com/weltevrede/psrsalsa} (A Suite of ALgorithms for
Statistical Analysis of pulsar data), are described in some detail.

As reported by \cite{cl86}, PSR B1839--04 (J1842--0359) was discovered
with the Lovell telescope at Jodrell Bank. It is a slow rotator
($P=1.84$ s; \citealt{hlk+04}), with a relatively large
characteristic age ($\tau_c=57$ Myr), which places it somewhat
close to the radio death line (e.g. \citealt{cr93a}). Nevertheless,
its spin properties do not appear to be in any way remarkable.

Most of the existing literature about PSR B1839--04 focuses on the
time-averaged radio properties of this pulsar. For instance,
\cite{wmlq93} reported that at 1560 MHz the emission is moderately
linearly polarized (25\%). The double-peaked pulse profile is
relatively wide, with the width measured at 10\% of the peak intensity,
$w_{10}$, being $66\degr$. This, in combination with the observed
polarization position angle as a function of rotational phase, led
these authors to suggest that the magnetic inclination is very small
($\alpha=9\degr$). These results are consistent with other studies
(e.g. \citealt{gl98,wmlq93,whx02,cw14}).

\cite{wes06} studied a large sample of radio pulsars for individual
pulse modulation. They found that the modulation of PSR B1839--04
shows a high degree of organisation with clearly defined drifting
subpulses at a centre frequency of 1380 MHz. This means that in a plot
of the intensity as a function of pulse number and rotational phase a
repeating pattern can be discerned that is reminiscent of an imprint
left by tires on a soft surface (see Fig. \ref{FigPulseStack}, which
is described in more detail below). The systematic shift in
rotational phase of these subpulses, that is, the substructure
discernible in the shape of an individual pulse, is often ascribed to
physics in the charge depletion region directly above the magnetic
pole (the polar gap). In such a scenario the polar gap produces radio
beamlets associated with ``sparks'', which are distributed on a
circular path around the magnetic axis. In such a ``carousel model''
this pattern of beamlets circulates around the magnetic axis because
of a so-called $\bm{E}\times\bm{B}$ drift, giving rise to the observed
pattern of drifting subpulses (e.g. \citealt{rs75,gmg03}).

That drifting subpulses were discovered for PSR B1839--04 was
not remarkable in itself, since they exist in at least one third of the
pulsar population \citep{wes06}. What makes this pulsar highly unusual
is that the drifting subpulses detected in the two components of the
pulse profile appear to drift in opposite directions, a phenomenon
dubbed bi-drifting by \cite{qlz+04}. Bi-drifting has only been
observed for one other pulsar: \object{PSR J0815+0939} \citep{mlc+04,clm+05}.
The pulse profile of PSR J0815+0939 at 430 and 327 MHz shows four very
distinct components. The leading component has an unclear drift sense
that appears to change during observations. The following component
shows subpulses drifting towards later phases (positive drift), while
negative drifting is observed for the remaining two components.

Bi-drifting is hard to interpret in the framework of circulating
beamlets. Two models have been put forward to explain bi-drifting, but we argue in Sect. \ref{SectDiscussion} that they both have a
fundamental problem in explaining the extremely high degree of
coherence in the modulation cycles observed for the regions showing
positive and negative drift in PSR B1839--04. The first model is based
on an ``inner annular gap'' \citep{qlw+04}, the second model
includes both proton and ion acceleration \citep{jon13,jon14}. Both
models allow for circulation to occur in opposite directions in
different parts of the gap region above the magnetic pole.

The manuscript is organised as follows. In the next section the
functionality of the PSRSALSA package is briefly summarised. Section
\ref{SectObs} describes the observations of PSR B1839--04, followed by
the results of the subpulse modulation analysis and a description of two
emission modes that were not reported before in the literature
(Sect. \ref{SectModulation}). The polarization properties are
analysed in Sect. \ref{SectPolarization} and the consequences of
these new results for existing models are discussed in Sect.
\ref{SectDiscussion}, followed by a summary in Sect.
\ref{SectConclusions}.

\section{PSRSALSA}
\label{SectPSRSALSA}

The analysis of pulsar data requires specialised software to
make progress in understanding these exciting and extreme
plasma-physics laboratories. Over the years the author has developed a
number of different tools to analyse pulsar data in various
ways. These tools have some overlap with tools provided by for
instance  PSRCHIVE\footnote{http://psrchive.sourceforge.net/}
\citep{hvm04}, but the emphasis is on single-pulse modulation and
polarization analysis. The functionality of the PSRSALSA package includes the following:
\begin{itemize}
\item Tools to analyse periodic subpulse modulation in various ways \citep{wes06,wse07,wwj12,ssw09}.
\item A tool to fit the rotating vector model (RVM; \citealt{rc69a}) to the polarization position angle curve of pulsars to derive viewing geometries \citep{rwj15}.
\item Tools to analyse and fit the observed flux-distribution of 
individual pulses, including the possibility to remove the effect of the noise distribution during the fitting process (e.g. \citealt{wws+06}).
\item Plotting tools for various types of data, visualised in various different ways, either from the command line or interactively.
\item Functionality to do various standard data processing operations, including de-dispersion, averaging, rebinning, rotating data in pulse phase and removing the average noise level (baseline) in various ways. There is functionality to flag and remove frequency channels and subintegrations from a data set.
\item Support of the PSRFITS \citep{hvm04}, SIGPROC\footnote{http://sigproc.sourceforge.net/} \citep{lor11}, EPN \citep{ljs+98}, and ASCII data formats.
\end{itemize}
PSRSALSA makes use of slalib for astronomical
calculations\footnote{Slalib is part of Starlink:
  http://www.starlink.ac.uk}, the GNU Scientific
Library\footnote{http://www.gnu.org/software/gsl/}, FFTW
3\footnote{Fastest Fourier Transform in the West:
  http://www.fftw.org/},
PGPLOT\footnote{http://www.astro.caltech.edu/~tjp/pgplot/}, and
cfitsio\footnote{http://heasarc.gsfc.nasa.gov/fitsio/}.  By describing
the analysis of the PSR B1839--04 data, many of the main features of
PSRSALSA are introduced and used. It is expected that more
features will be added in the future.

\section{Observations}
\label{SectObs}

\begin{table}[!tb]
\caption{\label{TableObs}Observational parameters of the analysed data.}
\begin{center}
\begin{tabular}{c|ccccc}
\hline
\hline
ID & MJD & Freq. & BW    & $t_\mathrm{samp}$ & $N_\mathrm{pulses}$\\
   &     & [MHz] & [MHz] & [msec]\\
\hline
2003  & 53004.4 & 1380 & 80 & 9.830 & 1025\\
2005a & 53674.6 & 1380 & 80 & 0.819 & 6008\\
2005b & 53677.8 & 1380 & 80 & 0.819 & 2162\\
2005c & 53686.8 & 1380 & 80 & 0.819 & 4378\\
\hline
\end{tabular}
\tablefoot{The first column defines the identifier used in the text, followed by the MJD, centre frequency, bandwidth, sampling time, and the number of recorded stellar rotations.}
\end{center}
\end{table}

PSR B1839--04 was observed once in 2003 and three times in 2005 with
the Westerbork Synthesis Radio Telescope (WSRT). An explanation of the
process of recording the data and the generation of a pulse-stack, an
array of intensities as a function of rotational phase and pulse number,
can be found in \cite{wes06}. The most relevant parameters related to
these observations can be found in Table \ref{TableObs}, including the
definitions of identifiers used in the text to refer to the individual
observations. Observation 2003 is the observation analysed in
\cite{wes06}.

Before proceeding with the analysis, it is important to subtract the
average noise level (baseline) from the data. This is complicated
because the baseline is time dependent. As long as the timescale of the
fluctuations is long compared to the pulse period, removing
the baseline should not be problematic. The level of the baseline can
be determined by considering the noise values in the off-pulse region
(those pulse longitudes where no pulsar signal is detected). PSRSALSA
offers the possibility to remove a running mean, which can be
especially useful if the number of off-pulse bins is limited. However,
for these data sets the baseline was independently determined for each pulse by calculating the average intensity of the off-pulse
bins. These baseline-subtracted data sets were used in the analysis
described in the next section.

\section{Subpulse modulation and emission modes}
\label{SectModulation}

\subsection{Basic appearance of the emission variability}

\begin{figure}
\begin{center}
\includegraphics[height=\hsize,angle=270]{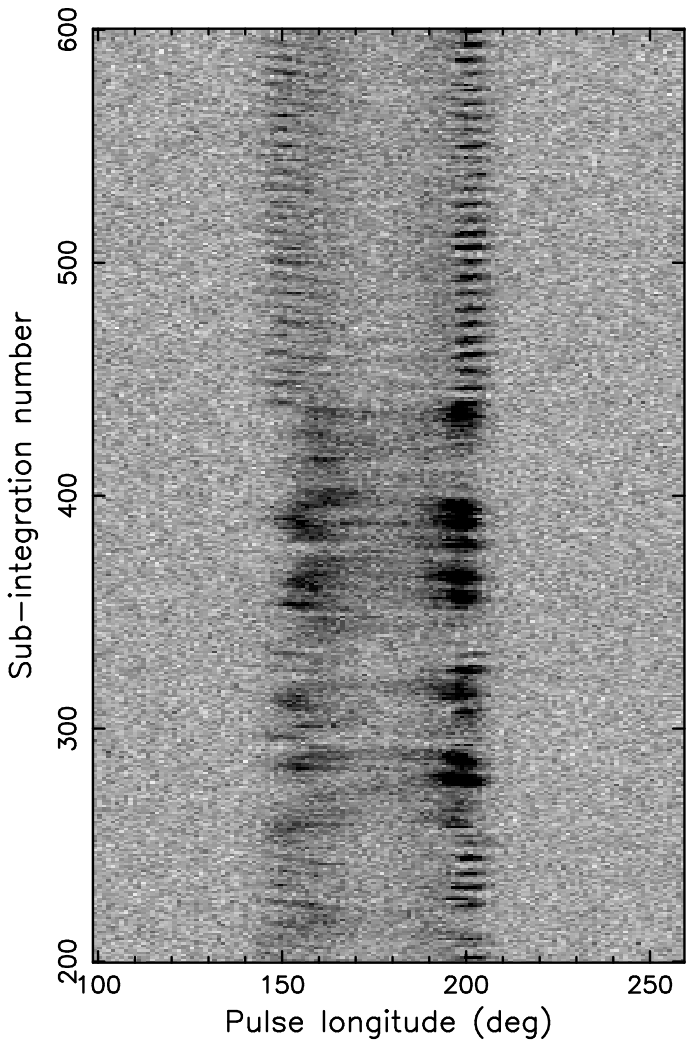}
\end{center}
\caption{\label{FigPulseStack}Pulse stack of PSR B1839--04. This is
  data set 2005a, rebinned into 300 pulse longitude bins. Every two
  pulses are summed together, so each subintegration contains the
  signal of two rotational periods. The data are clipped, meaning
that all
  intensities above a certain threshold are set to that threshold
  value, thereby restricting the dynamic range. This makes the weaker features better visible.}
\end{figure}

PSRSALSA can visualise the data in various ways. In
Fig. \ref{FigPulseStack} a pulse stack is shown in grey-scale. The
data shown in this figure are modified in several ways to emphasise
the weaker emission features. To increase the signal-to-noise
($S/N$) ratio of each sample, the data were rebinned from 2246 to 300
pulse longitude bins\footnote{Within PSRSALSA it is not required that
  $2^n$ samples, or even an integer number of samples, are summed to form a new pulse longitude bin. If, for example, the resolution is reduced by a
  factor 2.5, the first sample written out will be the sum of the
  first two samples plus half the value of the third sample.}.
Secondly, since in the case of this pulsar the pulse shapes do not
change too rapidly from pulse to pulse, each pair of pulses was
summed, so that the first row in the figure is the sum of the first two
pulses. These two steps greatly increase the $S/N$ per plotted
intensity sample, hence improve the visibility of the pulse shape evolution. Finally,
to make the organised patterns in the pulse stack even clearer, the data
were clipped. This means that intensity values above a certain
threshold were set to the threshold value. As a consequence, black
regions in Fig. \ref{FigPulseStack} which appear to have a uniform
intensity, might in reality show some intensity variation. However,
since the colour scale varies more rapidly over a narrower
intensity range, intensity variations in weaker emission become much
clearer.

Figure \ref{FigPulseStack} shows several properties of the emission
that are analysed in more detail later in this section. First of
all, the emission is separated into a leading and trailing pulse
longitude component that are referred to as components I
($\sim140\degr-170\degr$) and II ($\sim180\degr-210\degr$). These main
profile components will be shown to be composed of two largely overlapping components themselves, resulting in components Ia, Ib, IIa, and IIb. Throughout this
manuscript, including Fig. \ref{FigPulseStack}, the pulse longitude
axis is defined such that component II peaks at $200\degr$, hence
placing the saddle region between the two peaks approximately at
$180\degr$.

The pulse stack shows that the emission of PSR B1839--04 is highly
variable. Firstly, a clear periodic on- and off-switching is visible, which
is clearest at the top of the figure from subintegration 450 and
above. The bright emission phases are separated by approximately six
subintegrations, hence 12 stellar rotations. This phenomenon is known
as drifting subpulses, and the repetition period is referred to as
$P_3$. Especially for component I, centred at pulse longitude
$\sim150\degr$, the bright emission phases appear as bands that are
slightly diagonal in the figure, indicating that the bright emission
``drifts'' toward earlier times in subsequent stellar rotations. The
emission bands seen in the trailing half of the profile are
essentially horizontal with little evidence of drift, although for
some bands a small tilt in the opposite direction compared to those in
component I can be observed.  These effects are quantified in
Sects. \ref{SectFluctuation} and \ref{SectSubpulsePhase}.

Secondly, Fig. \ref{FigPulseStack} shows that the subpulse behavior is
distinctly different between subintegrations $\sim250$ and $\sim450$,
indicating that two distinct emission modes operate in this
pulsar. The mode referred to as the quiet mode (or Q mode) shows the
regular and highly periodic subpulse modulation. However, during the
more intense bright mode (B mode) the periodic subpulse modulation
appears to be far less pronounced, or even absent. During the B mode
the emission is not only brighter, components I and II are also closer
to each other in pulse longitude with more emission in the saddle
region. The modes are described in more detail in the next
section.

\subsection{Mode separation and concatenating data sets}
\label{SectModeSep}

To quantify the difference in the emission properties of the
B and Q mode, the start and end pulse number of each mode was identified. Because of the variability caused by
subpulse modulation, this process cannot easily be automated
reliably. Instead, the classification was made by eye by
interactively plotting the pulse stack using PSRSALSA, allowing
smaller subsets of the data to be investigated
efficiently. Identifying the mode transitions for this pulsar is a somewhat
subjective process because of the intrinsic variability observed
during a mode. Therefore it is often not clear with a single-pulse
precision when a mode starts or ends. For instance, in
Fig. \ref{FigPulseStack} it might be argued that the B mode is interrupted
for a short Q mode around subintegration 330, but it is hard to be
certain. In addition, the transitions sometimes appear to be somewhat
gradual, which could be because of the limited $S/N$ of the
single pulses. To ensure that as much of the data as possible is
labelled with the correct mode, some of the data were not assigned to
either mode.

For all individual observations the fraction of rotations without an
assigned mode was less than 10\% (the average was 3\%), except for the
2005c observation, for which 35\% was in an
undetermined mode. This was because the source was setting towards the
end of the observation, resulting in an increase of the system
temperature. Ignoring this observation, of the remaining 9195
rotations, 25\% and 72\% were assigned to the B and Q mode,
respectively, with the remaining 3\% being unclear. The median duration of a B and Q
mode was 85 and 270 pulse periods, respectively.

To improve the $S/N$ of the following analysis,
it is desirable to combine the data sets. PSRSALSA allows different
observations to be concatenated, which was only done for the three
2005 data sets, since they have the same time resolution. The
data sets were aligned in pulse longitude in a two-step
process. First, an analytic description of the 2005a profile was
obtained by decomposing the profile in von-Mises functions using
PSRCHIVE. This allows PSRSALSA to align the data sets by
cross-correlating this analytic description of the profile with the
remaining two data sets. Since the profile is different during  B
and Q modes, the pulse profiles obtained by summing all pulses will
vary from observation to observation. Therefore, the analytic
template was based on the 2005a Q-mode pulses alone. The offsets in
pulse longitude between the different data sets were determined by
only cross-correlating the template with the Q-mode emission of all observations. The offsets
obtained were applied to the full observations before concatenation.

During the process of concatenation, it is possible to specify that
only multiples of a given block size defined in number of pulses are
used from each data set. Similarly, when the mode separation was
applied, it was possible to only use continuous stretches of data of
the same specified length. This results in a combined data set that
consists of blocks of continuous stretches of data of a given length,
which is important when analysing fluctuation properties of the data
with Fourier-based techniques (see Sect. \ref{SectFluctuation}).

\begin{figure}
\begin{center}
\includegraphics[height=\hsize,angle=270,trim=15 0 0 0,clip]{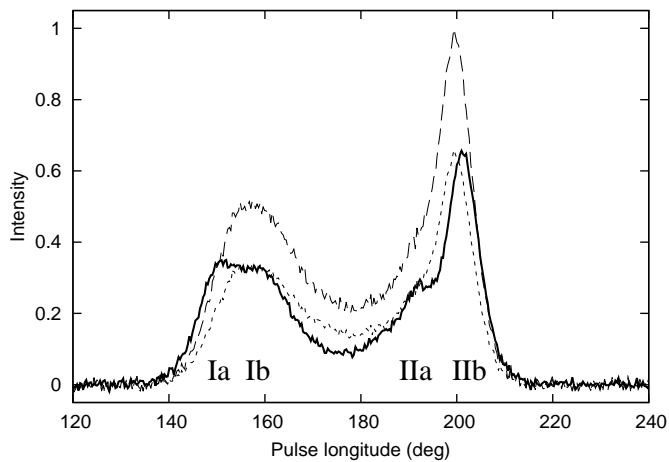}
\end{center}
\caption{\label{FigModeProfiles}Pulse profile of the different
  modes of PSR B1839--04. These are the combined 2005 data sets (the Q-
  and B-mode profiles consist of 7605 and 3188 individual pulses,
  respectively), rebinned with a factor two. The thick solid line and
  the long-dashed profile show the Q- and B-mode profiles,
  respectively. The peak flux of the B-mode profile is normalised and
  the Q-mode profile is scaled accordingly after taking into account
  the difference in number of pulses. The dotted profile is a
  scaled-down B-mode profile to aid a comparison between the shapes of
  the profiles in the two modes. The location of the four profile
  components are labelled.}
\end{figure}

Figure \ref{FigModeProfiles} shows the pulse profiles of the two
modes. As expected, the B mode is brighter than the Q mode (compare
the highest amplitude profile with the thick solid line). In addition,
the saddle region of the profile is more pronounced
in the B mode (compare the thick solid line with the dotted line,
which is a scaled-down B-mode profile). As noted from an inspection of
the pulse stack (Fig. \ref{FigPulseStack}), the two components are
slightly closer together in the B mode. Figure \ref{FigModeProfiles}
demonstrates that this is predominantly because the leading component
shifts to later phases, although the trailing component also shifts
slightly inwards. Especially the Q-mode profile shows that the main
profile components show structure. In Fig. \ref{FigModeProfiles} four
profile components are defined and labelled.

\subsection{Pulse energy distribution}

\begin{figure}
\begin{center}
\includegraphics[height=0.99\hsize,angle=270,trim=165 7 0 234,clip]{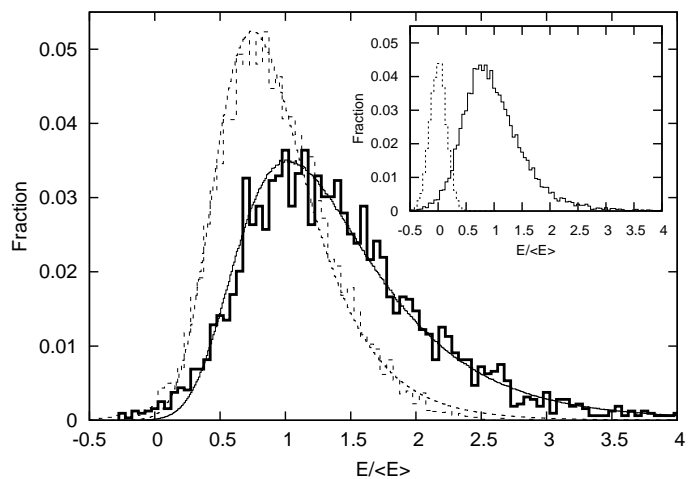}
\end{center}
\caption{\label{FigEnergyDist}{\em Inset:} The normalised pulse energy distribution of the combined 2005 data sets (solid histogram), and that of the off-pulse region (dotted histogram). The latter is scaled down by a factor three to make their amplitudes comparable.
{\em Main panel:} The similarly normalised energy distribution of the Q (thin dashed histogram) and B mode (thick solid distribution) are shown separately. The fitted distribution functions are shown as the thin solid (B mode) and dashed curve (Q mode).}
\end{figure}

The difference in the brightness of the pulses in the two modes can be
further quantified by analysing the so-called pulse energy
distribution. The methods used here largely follow those
described in \cite{wws+06}. The pulse energies were calculated for
each individual pulse by summing the intensity values of the
pulse longitude bins corresponding to the on-pulse region. The pulse
energy distribution is shown in the inset of Fig. \ref{FigEnergyDist}
together with an energy distribution obtained by summing
off-pulse bins using an equally wide pulse longitude interval. Since the observations were not flux
calibrated, the obtained pulse energies were normalised by the average
pulse energy $\left<E\right>$. 

The pulse energy distribution including both modes shown in the inset
of Fig. \ref{FigEnergyDist} shows no clear evidence for
bi-modality. Bi-modality could have been expected since the
distribution contains a mixture of bright and quiet mode pulses which,
per definition, should have a different average brightness. The
absence of bi-modality is a consequence of the significant
pulse-to-pulse brightness variability observed during a mode, although
random noise fluctuations can be expected to broaden the intrinsic
energy distributions as well. The latter is quantified by the
off-pulse distribution shown as the dotted histogram in the inset. The
absence of clear bi-modality makes it hard to separate the modes
in a more objective way than the method described in
Sect. \ref{SectModeSep}.

The energy distributions of the two modes are shown separately in the
main panel of Fig. \ref{FigEnergyDist}. The B-mode pulses are
indeed brighter on average than the Q-mode pulses. As expected,
the distributions significantly
overlap. The cross-over
point between the two distributions is located close to the average pulse
energy $\left<E\right>$, which is where, in the overall energy
distribution shown in the inset of Fig. \ref{FigEnergyDist}, a small
kink can be identified.

The distributions of both modes can be fairly well characterised by an
intrinsic lognormal distribution convolved with the noise
distribution. The lognormal distribution is defined to be
\begin{align}
P(E/\left<E\right>) = \frac{\left<E\right>}{\sqrt{2\pi}\sigma E}\exp\left[-\left(\ln \frac{E}{\left<E\right>} -\mu\right)^2/\left(2\sigma^2\right)\right].
\end{align}
The parameters $\mu$ and $\sigma$ were refined in PSRSALSA using the
Nelder-Mead algorithm\footnote{The PSRSALSA code is based on an
  implementation by Michael F. Hutt;
  http://www.mikehutt.com/neldermead.html} \citep{nm65}, which is an
iterative process. For each iteration a large number of values of
$E/\left<E\right>$ were drawn at random from the lognormal
distribution. A randomly picked value from the off-pulse distribution
was added to each value, resulting in a lognormal distribution
convolved with the noise distribution. The goodness of fit used during
the optimisation is based on a $\chi^2$ test using the cumulative
distribution functions of the model and observed distribution and does
therefore not depend on an arbitrarily chosen binning of the
distributions. More detailed information about the method is available
in the help of the relevant tools\footnote{Named {\sc pdistFit} and {\sc pstat}.} in PSRSALSA.

\begin{table}[!tb]
\caption{\label{TablePenergyFit}Parameters quantifying the observed energy distribution of the Q and B modes after correcting for the effect of the noise distribution. }
\begin{center}
\begin{tabular}{c|cccc}
\hline
\hline
Mode & $\mu$ & $\sigma$ & KS prob. & Rejection significance\\
\hline
B &  0.23 & 0.49 & $1.4\times10^{-1}$ & 1.5\\
Q & $-0.15$ & 0.44 & $2.7\times10^{-3}$ & 3.0\\
\hline
\end{tabular}
\tablefoot{The parameters $\mu$ and $\sigma$ define a lognormal distribution. The next column gives the probability derived from a KS-test between the observed and modelled distribution, followed by the significance level (in standard deviations of a normal distribution) that the null hypothesis can be rejected, 
which says that the samples are drawn from the same distribution.}
\end{center}
\end{table}

Table \ref{TablePenergyFit} lists the results of the fitting
process. The overall goodness of fit after optimisation was quantified
using the Kolmogorov-Smirnov (KS) test implemented in PSRSALSA (see
e.g. \citealt{ptv+92} for a description of a similar implementation of
the algorithm). It is a non-parametric test of the model and observed
distribution, which does not require binning either. The null hypothesis
is that the samples from both distributions are drawn from the same
parent distribution. The probability that random sampling is
responsible for the difference between the model and observed
distributions can be estimated (hence a small probability means that
the distributions are drawn from a different distribution). This
probability can be expressed as a significance, in terms of a standard
deviation of a normal distribution, of the rejection of the null
hypothesis. In other words, a high significance means that the two
distributions are likely to be different\footnote{A low significance
  does not imply that the distributions are drawn from the same parent
  distributions. It only quantifies that there is no evidence that the
  distributions are different.}. The probability and significance are
quoted in Table \ref{TablePenergyFit}. Although
the B mode is described with sufficient accuracy by a lognormal
distribution, the fit of the Q-mode distribution is less accurate.

These results are consistent with the results of \cite{bjb+12}, who
also used a lognormal distribution to describe the observed energy
distribution. A detailed comparison is not possible because \cite{bjb+12} neither separated  the emission into different
modes, nor did they convolve the model distribution with the noise
distribution to obtain the intrinsic distribution. Nevertheless, their
parameters are in between those derived for the B and Q modes in this
work.

\subsection{Fluctuation analysis}
\label{SectFluctuation}

A sensitive method to determine time-averaged properties of the
periodic subpulse modulation is the analysis of fluctuation
spectra. The techniques described here are largely based on the work
of \cite{es02}, and for more details about the analysis, we also refer
to \cite{wes06,wse07,wwj12}. Two types of fluctuation spectra are
exploited here, the first being the longitude-resolved fluctuation
spectrum (LRFS; \citealt{bac70b}). The LRFS is computed by separating
the pulse stack into blocks with a fixed length. As explained in more
detail below, in this case, block lengths of either 512 or 128 pulses
were used. For each block and each column of constant pulse longitude, a
discrete Fourier transform (DFT) was computed, revealing the presence
of periodicities for each pulse longitude bin. The final LRFS is
obtained by averaging the square of the spectra obtained for the
consecutive blocks of data.

\begin{figure*}
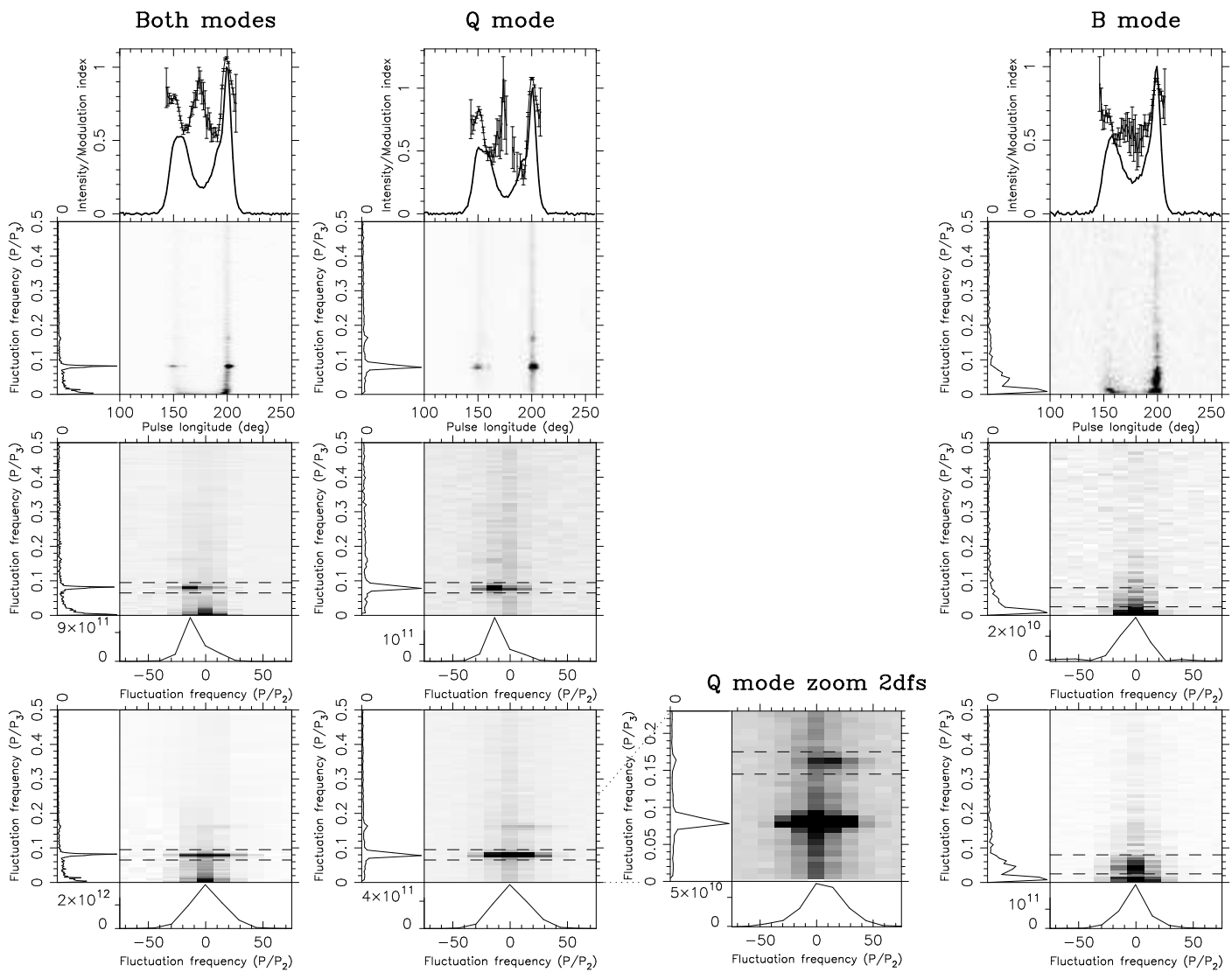

\begin{center}
\includegraphics[height=0.24\hsize,angle=270]{spectra_all2005.ps}
\hspace{0.0\hsize}
\includegraphics[height=0.24\hsize,angle=270]{spectra_Qremoved2005.ps}
\hspace{-0.015\hsize}
\begin{rotate}{270}
\hspace{0.575\hsize}
\begin{tikzpicture}
\draw[dotted] (0.017\hsize,0.0\hsize) -- (-0.05\hsize,0.057\hsize);
\draw[dotted] (0.089\hsize,0.0\hsize) -- (0.089\hsize,0.035\hsize);
\end{tikzpicture}
\end{rotate}
\hspace{0.01\hsize}
\includegraphics[height=0.25\hsize,angle=270,trim=-328 0 0 0,clip=true]{spectra_Qzoom2005.ps}
\hspace{0.0\hsize}
\includegraphics[height=0.24\hsize,angle=270]{spectra_Bremoved2005.ps}
\end{center}
\caption{\label{FigSpectra}Fluctuation analysis of the emission
  including both modes of PSR B1839--04, and those of the Q and
  B mode separately. These are the combined 2005 data sets, rebinned
  with a factor 9 in pulse longitude. {\em Top row:} The pulse
  profile and the modulation index (points with error bars). {\em
    Second row:} The LRFS and a side panel showing the horizontally
  integrated power. {\em Third row:} The 2DFS of component I and side
  panels showing the horizontally (left) and vertically, between the
  dashed lines, integrated power (bottom). {\em Bottom row:} The 2DFS of
  component II. {\em Third column:} The 2DFS of Q-mode emission of
  component II with an adjusted vertical range and grey-scale
  emphasizing the second harmonic.}
\end{figure*}

The fluctuation analysis was made for the Q and B modes
separately, and for both modes combined. To achieve the highest $S/N$,
the combined 2005 data sets were used. However, it is important to
ensure that the combined data set consists of an integer number of
blocks with an equal number of single pulses $n_\mathrm{dft}$ each,
where $n_\mathrm{dft}$ is the length of the DFTs used to compute the
fluctuation spectra. This ensures that the single pulses used in each
DFT are continuous. For the left column of Fig. \ref{FigSpectra},
which shows the results of the fluctuation analysis before mode
separation, $n_\mathrm{dft}$ was set to 512 pulses to achieve a
relatively high frequency resolution. Since this implies that
only a
multiple of 512 pulses could be used, some of the recorded pulses of
each 2005 data set were not included in the analysis (resulting in
11776 analysed pulses). The fluctuation analysis of the individual
modes was hampered by the fact that the mode changes occur too often
to find long enough stretches of data without a mode switch. Therefore
$n_\mathrm{dft}$ was reduced to 128 consecutive pulses for the mode-separated data.
Next, some general results for the non-mode-separated
data are summarised and the methods used are explained. This is followed
by the results of the mode-separated data.

\subsubsection{Non-mode-separated data}

In Fig. \ref{FigSpectra} the top panels show the pulse profile (solid
curve). The panel below shows the LRFS in grey-scale, which is aligned in pulse
longitude with the panel above. The side panel of the LRFS
of the whole data set (first column) shows a
clear spectral feature at $P/P_3\simeq0.08$ cycles per period
(cpp). This spectral feature corresponds to the pattern repetition
period of the drifting subpulses $P_3\simeq12P$ that was identified by eye in the pulse stack
(Fig. \ref{FigPulseStack}). The narrowness of the spectral feature
(high quality factor) indicates that the periodicity is well
defined. The LRFS itself shows that most of this periodic modulation power is
associated with the outer halves of the two main components where most of the spectral power is detected (darker colour),
that is, components Ia and IIb corresponding to pulse longitudes
$145\degr-155\degr$ and $195\degr-205\degr$. As mentioned before and illustrated further in Sect. \ref{SectSubpulsePhase}, this
suggests that the two main profile components are double themselves,
resulting in at least four distinct profile components. Although much weaker, the second harmonic at $\sim0.16$ cpp is
detected as well.

In addition to this well-defined spectral feature, a weak
white noise component is most clearly seen as the vertical darker band in
the LRFS centred at pulse longitude $\sim200\degr$. This indicates
non-periodic random fluctuations. A red-noise-like component is strongest below $P/P_3=0.02$
cpp, corresponding to fluctuations on a timescale of 50 pulse periods
and above.

The power in the LRFS can be used to quantify the longitude-resolved
modulation index, which is shown as the points with error bars in the
top panels of Fig. \ref{FigSpectra} together with the pulse profile. This is a measure for the amount
of intensity variability (standard deviation), normalised by the
average intensity at a given pulse longitude. An advantage of
computing the modulation index through the spectral domain is that for
instance periodic radio frequency interference (RFI), which affects a specific spectral channel, can be
masked, thereby removing its effect on the modulation index. In addition, the
effect of white noise as determined from the off-pulse region,
resulting from a finite system temperature, for
instance, can be
subtracted from the measured modulation index.

PSRSALSA can determine the error bar on the modulation index using a
method analogous to that used by \cite{es02}, or through bootstrapping,
which generally results in more reliable (although conservative)
errors (see \citealt{wwj12} for details). A pure sinusoidal drifting
subpulse signal would result in a modulation index of
$1/\sqrt{2}\simeq0.71$.  Since both lower and higher values are
observed at different pulse longitude ranges (see
Fig. \ref{FigSpectra}), the presence of both non-fluctuating power and
additional non-periodic intensity fluctuations can be
inferred. \cite{bjb+12} found the minimum in the longitude-resolved
modulation index of PSR B1839--04 to be $1.12$, significantly higher
than what is shown in Fig. \ref{FigSpectra}. This is most likely
because the authors included system-temperature-induced white noise in
the result (private communication).

The LRFS is a powerful tool to identify any periodically repeating
patterns as a function of pulse longitude, but it cannot be used to
identify if the emission drifts in pulse longitude from pulse to pulse
(i.e. drifting subpulses). If drifting subpulses are present, the
pattern of the emission in the pulse stack resembles a two-dimensional
sinusoid. This makes the two-dimensional fluctuation spectrum (2DFS;
\citealt{es02}) especially effective in detecting this type of pattern
since most modulation power is concentrated in a small region
in this spectrum.

The 2DFS was computed for components I and II separately,
resulting in the grey-scale panels in the one but last and last rows in
Fig. \ref{FigSpectra}, respectively. Like for the LRFS, the vertical
frequency axis of the 2DFS corresponds to $P/P_3$, and the spectra for
both components show the same $P_3\simeq12P$ pattern repetition periodicity as identified
in the LRFS. The horizontal axis of the 2DFS denotes the pattern
repetition frequency along the pulse longitude axis, expressed as
$P/P_2$, where $P_2$ corresponds to the separation between the
subpulses of successive drift bands in pulse longitude, that is, their horizontal separation in Fig. \ref{FigPulseStack}. The first 2DFS in the left
column of Fig. \ref{FigSpectra} shows that the power corresponding to the
drifting subpulses (found at $P/P_3\simeq0.08$ cpp between the dashed lines) is asymmetric, such that most of the power is
concentrated in the left-hand half of the spectrum. This indicates
that the subpulses move towards earlier pulse longitudes from pulse to
pulse. The power peaks at $P/P_2\simeq-15$ cpp, suggesting that
$P_2\simeq360\degr/15\simeq25\degr$. 
This implies that the drift bands in component I have a drift rate (slope in the pulse stack) $P_2/P_3\simeq2\degr$ per $P$ and thus that the subpulses drift.
The 2DFS of the trailing
component shows again clear $P_3$ modulation, but the spectral feature
is much more centred on the vertical axis. This suggests that if the
subpulses do drift in pulse longitude, $P_2$ will be much larger for
the trailing component, corresponding to more horizontal drift bands
in the pulse stack. This is consistent with Fig. \ref{FigPulseStack}, which showed little evidence for drift in component II.

\subsubsection{Mode-separated data}
\label{SectFluctModeSep}

The second and last column of Fig. \ref{FigSpectra} show the results
of the same type of analysis, but now for the mode-separated data. As
discussed before, it is important to use blocks of continuous
stretches of data with a length determined by $n_\mathrm{dft}$. Since
the modes only last for a certain amount of time, the amount of
usable data decreases the larger $n_\mathrm{dft}$ is. For this reason
a reduced $n_\mathrm{dft}$ of 128 pulses was used for the mode
separated data.  The Q- and B-mode analysis is based on all available
continuous stretches of 128 pulses in a given mode found in each of the
2005 data sets, resulting in 6016 and 896 available pulses for the Q
and B mode, respectively.
The spectra of the Q mode clearly show the $P_3\simeq12P$ periodic
subpulse modulation, while this feature is absent from the B-mode data. On the other hand, the B mode shows a low-frequency red-noise
component that is absent during the Q mode. This clearly
indicates that the features in the spectra seen in the first column
of Fig. \ref{FigSpectra} are the superposition of features seen in the
two modes.

\cite{wes06} measured $P_2$ and $P_3$ from the 2DFS by calculating
the centroid of a region in the 2DFS containing the drifting subpulse
feature, which therefore does not rely on any assumption about the
shape of the feature. The associated uncertainties are typically the
combination of three effects. First of all, there is a statistical
error caused by white noise in the data (hence in the 2DFS), resulting
in an uncertainty in the centroid calculation. Typically, this error
can be neglected because there is a larger systematic error
associated with the somewhat subjective choice of which rectangular
region in the 2DFS to include in the centroid calculation. This error
can be estimated in PSRSALSA by using an interactive tool that allows
the user to select a rectangular region in the 2DFS after which the
centroid and statistical error is determined. By repeating this
process a number of times by making different selections, this
systematic error can be quantified. The third contribution to the
uncertainty is that pulsar emission is in practice inherently erratic
in the sense that if a certain feature is statistically significant
for a certain data set, it still might not be an intrinsic
(reproducible) feature of the pulsar emission. This is always an issue
when analysing a limited data set that is too short to determine the
true average properties of the pulsar emission. To assess the
importance of this effect, the data set could be split in different
ways to see if the same feature is persistent. In addition, the order
of the pulses can be randomized to determine the strength of any
apparent ``significant'' periodic features, which in this case clearly
must be the result of having a finite data set. For PSR
B1839--04, the features reported here are confirmed in all data
sets,
ensuring that they truly reflect intrinsic properties of the pulsar
emission.

After taking the above systematics into consideration, \cite{wes06}
reported (based on the 2003 observation) that components I and II show
periodic modulation with $P_3=12.4\pm0.3\;P$ and $P_2=-35^{+5}_{-15}$
and $120^{+20}_{-20}$ degrees, respectively. Note in particular the
sign difference in the two $P_2$ values, implying that the subpulses
drift towards earlier phases in component I, but in the opposite
direction for component II. This is referred to as bi-drifting
and the existence of this phenomenon has important theoretical
consequences. The claim for bi-drifting was based on a lower $S/N$
equivalent of the plot in the first column (not mode-separated data) of
Fig. \ref{FigSpectra}. The $P_2$ value of the leading component is
convincingly negative (see bottom side-panel of the first 2DFS, which
shows that the vertically integrated power between the two dashed
lines clearly peaks at a negative $P_2\simeq-15$ cpp. Bi-drifting should
result in a feature at the same $P_3$ in the second 2DFS (bottom row of Fig. \ref{FigSpectra}),
which is offset in the opposite direction. This is not immediately
obvious from the spectrum (even with the greatly increased $S/N$),
since the power peaks at 0 cpp, suggesting no drift at
all. Nevertheless, utilising the centroid method outlined above, it is
found that $P_3=12.4\pm0.2\;P$ and $P_2=-35^{+7}_{-5}$ degrees for
component I and $P_3=12.4\pm0.1\;P$ and $P_2=260^{+60}_{-10}$ degrees
for component II. These measurements are consistent with those by
\cite{wes06} and confirm the slight offset of the centroid of the
feature in the 2DFS of the second component, resulting in the claim of
bi-drifting. However, given the importance of the claim, and the level
of uncertainty in the errors, more convincing evidence is highly
desirable.

By exploiting the high $S/N$ of the combined 2005 data, further evidence
for bi-drifting comes from analysing the second harmonic in the
2DFS of component II.  This harmonic is shown more clearly in the
third column of Fig. \ref{FigSpectra}, which only shows the bottom
half of the 2DFS with the brightest features clipped to emphasize the
weaker features. This harmonic should be located twice as far from the
origin than the fundamental, that is, $P/P_3$ and $P/P_2$ can be
expected to be twice as large. Especially the increased offset from
the vertical axis helps greatly since the resolution in $P/P_2$ is
relatively poor\footnote{This resolution is set by the duty cycle of
the  pulse-longitude region used to calculate the 2DFS.}. As clearly
demonstrated by the bottom side-panel, the integrated power between
the dashed lines is asymmetric with an excess at positive
$P_2$, convincingly confirming bi-drifting. The centroid position of
the second harmonic was determined to be $P_3=6.1\pm0.2\;P$ and
$P_2=50^{+60}_{-2}$ degrees, consistent with the measurements based on
the fundamental. The separate 2005 data sets show a similar second
harmonic (not shown), and even the 2003 data analysed by \cite{wes06}
weakly show the same feature. Although the detection of a second
harmonic was mentioned by \cite{wes06}, its offset from the vertical
axis was neither measured nor used.

\subsection{Subpulse phase analysis}
\label{SectSubpulsePhase}

\begin{figure}
\begin{center}
\includegraphics[height=0.95\hsize,angle=270]{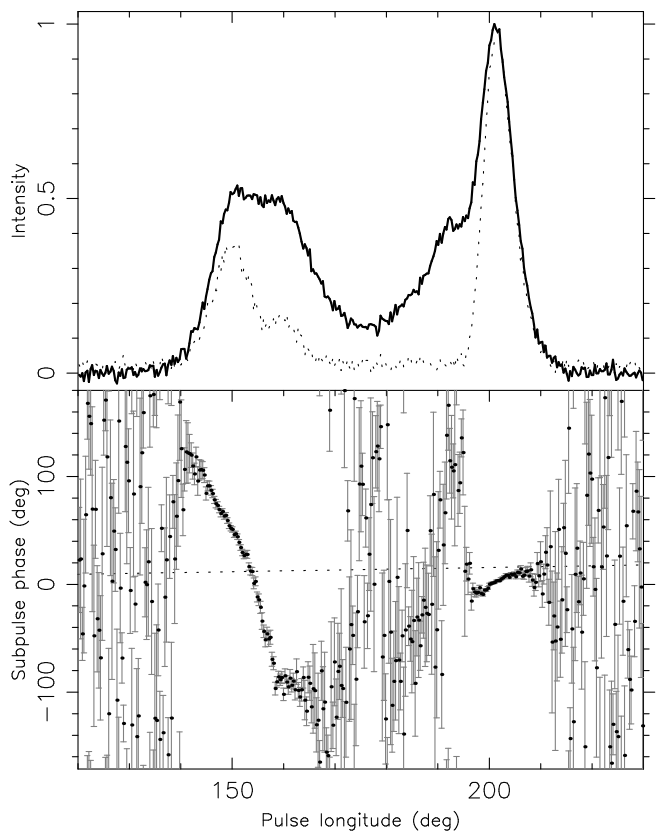}
\end{center}
\caption{\label{FigSubpulsePhaseTrack}These are the combined 2005 Q-mode
  data rebinned with a factor of 2. {\em Upper panel:} The strength of
  the $12P$ modulation pattern as a function of pulse longitude (dotted
  curve) and the pulse profile (black curve). {\em Lower panel:} The
  corresponding subpulse phase. An arbitrary phase offset has been
  applied to make the subpulse phase zero at the peak of the
  profile. The dotted line indicates the gradient of the phase
  relation expected for a modulation pattern that is longitude
  stationary in the co-rotating pulsar frame. }
\end{figure}

Although the 2DFS is very effective in detecting drifting subpulses in
noisy data, it does not allow easy characterisation of all the
relevant average properties. In particular, there is only limited
information about their pulse longitude dependence, including for
instance the phase relation of the modulation cycle as observed in
different components. Following \cite{es02} and as explained in
\cite{wwj12}, PSRSALSA allows this phase relation as a function of
pulse longitude to be extracted using spectral analysis. This
information, as shown in Fig. \ref{FigSubpulsePhaseTrack}, is obtained
by first of all computing the LRFS for the Q-mode data of the combined
data set. Relatively short DFTs were used: $n_\mathrm{dft}=64$
pulses. This ensured that the majority of the power associated with
the drifting subpulses was concentrated in a single spectral bin. In
addition, this allowed more blocks of consecutive pulses in the Q
mode
to be analysed: 6720 pulses in total.

The LRFS as discussed in Sect. \ref{SectFluctuation} was constructed
by taking the square of the complex values obtained from the DFTs,
while here the complex phase was considered as well. The complex phase
is related to the subpulse phase, while the magnitude of the
complex values is related to the strength of the drifting subpulse
signal. A difference in subpulse phase obtained for two pulse
longitude bins quantifies the difference in phase of the modulation
pattern at those two longitudes. In other words, if there is no phase
difference, the subpulse modulation pattern is in phase in the sense
that the maximum brightness occurs at the same pulse numbers (which is
often referred to as longitude stationary modulation).

Here the spectral bin centered at 0.078125 cpp was analysed, which
contains most of the subpulse modulation. The complex phase will be
different for the subsequent blocks of data because $P_3$ will be
variable over time\footnote{In addition, even if $P_3$ were
  constant, the complex phase will change from block
to
  block if it does not correspond to the centre frequency of the
  analysed spectral bin.}. As a consequence, a process is required to coherently add
the complex numbers obtained for the subsequent blocks of data. The
required phase offsets are independent of pulse longitude, but vary
from block to block. The iterative process used to obtain these phases
maximises the correlation between the phase profiles of the different
blocks in the on-pulse region. The coherent addition results in the
complex subpulse modulation envelope, with a complex phase
corresponding to the subpulse phase. Error bars were determined by
applying a similar method to that used to determine the uncertainties
on the modulation index. In addition to the complex phase of the resulting
complex modulation envelope, its amplitude is also obtained (see also
\citealt{es02,esv03}). The resulting subpulse amplitude envelope
as calculated with PSRSALSA is normalised such that it would be
identical to the normalised pulse profile if the subpulse modulation
were sinusoidal with a frequency $P/P_3$ equal to that of the
spectral bin analysed. The actual measured amplitude depends on the
presence of non-modulated power, the stability of the subpulse phase
differences between pulse longitude bins, and the accuracy of the
determined offsets used to do the coherent addition.

Figure \ref{FigSubpulsePhaseTrack} shows the results of this
analysis. For component I the subpulse phase decreases with
pulse longitude. In the sign convention used here (and in
\citealt{wws07,wwj12}), this corresponds to negative drift, that
is, drift
towards earlier pulse longitudes\footnote{Note that later pulse
  longitudes are observed later in their modulation cycle, so a
  decrease in subpulse phase as a function of pulse longitude is
  mathematically not necessarily the obvious choice of sign for
  negative drift. However, the chosen sign convention is preferred in
  PSRSALSA since the sign of the gradient of the subpulse phase track
  matches that of the appearance of drift bands in the
  pulse stack.}. The slope of the subpulse phase track is opposite for
the trailing component, confirming the bi-drifting
phenomenon\footnote{It must be noted that the results obtained from
  the 2DFS and the subpulse phase track are in some respects
  equivalent since they are both based on Fourier analysis, so this
  should not be considered to be completely independent confirmation
  of this result.}.

The top panel of Fig. \ref{FigSubpulsePhaseTrack} shows that the subpulse
amplitude profile is high in both the leading part of the leading
half of the profile (component Ia) and especially the trailing part
of the trailing half of the profile (component IIb). The subpulse
amplitude profile shows component Ib as a more distinct component
compared to the intensity profile. This again suggests that the pulse
profile should be considered to be composed of four components:
two pairs of components that are blended to form the roughly double-peaked pulse profile as indicated in Fig. \ref{FigModeProfiles}. This
could be interpreted as the line of sight making a relatively central
cut through a nested double-conal structure, with the modulation being
strongest in the outer cone.

Interestingly, the subpulse phase at the trailing end of component Ib
becomes longitude stationary, more like what is observed for component
IIb. In the bridge region between the two emission peaks the
error bars on the subpulse phase are large, partly because the
errors are somewhat conservative, but mainly because the emission barely shows the $12P$ modulation. Nevertheless, there appears to
be a clustering of subpulse phase points that smoothly connects the
subpulse phase tracks observed for the strongly modulated outer
components.  Possible relatively sharp deviations are observed at
pulse longitudes $\sim177\degr$ and $\sim192\degr$.

When interpreting the subpulse phase as a function of rotational phase, it 
should kept in mind that if the pulsar generates a modulation pattern
that is longitude stationary in the pulsar (co-rotational) frame,
that is, a pure intensity modulation without any drift, the observer will
still observe an apparent drift (e.g. \citealt{wws07}). This is because for a given pulse number, a later pulse longitude is
observed later in time, which means that a later part of the
modulation cycle will be observed. An intrinsically pure intensity
modulation will therefore result in a slow apparent positive drift
(i.e. emission appearing to drift toward later phases). This
drift rate is indicated by the dotted line in the lower panel of
Fig. \ref{FigSubpulsePhaseTrack}, which has a gradient of $P/P_3$
degrees of subpulse phase per degree of pulse longitude.

\subsection{Phase-locking}
\label{SectPhaseLocking}

As we discuss in Sect. \ref{SectDiscussion}, bi-drifting can be
more easily explained if the modulation cycle observed in the two
profile components operate, at some level, independent of each other. It
is therefore important to determine if the two modulation cycles are
phase locked, meaning that if the modulation cycle slows down in the
leading component, it is equally slow in the trailing component, such
that the modulation patterns of both components stay in
phase. Phase-locking can be expected for PSR B1839--04 given that the
subpulse amplitude in Fig. \ref{FigSubpulsePhaseTrack} is high for
both components. This would only occur if the subpulse phase
relationship with pulse longitude as shown in the lower panel is fixed
over time. Phase-locking is confirmed by computing and comparing the
subpulse phase profiles of the Q-mode data of the individual
observations listed in Table \ref{TableObs}. The shape of the subpulse
phase profile, including the phase delay between the components, is
the same for the separate observations (not shown), confirming that
strict phase-locking is maintained over a period of years. This
strongly suggests that the modulation patterns in both components have
a single physical origin.

\begin{figure}
\begin{center}
\includegraphics[height=\hsize,angle=270]{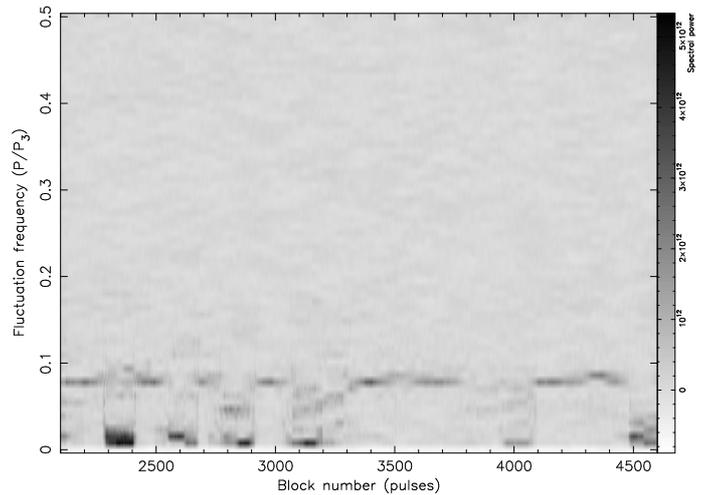}
\end{center}
\caption{\label{FigSlidingP3}Part of the S2DFS of observation 2005a of
  PSR B1839--04, showing spectral power as a function of pattern
  repetition frequency in component II and time (defined as the start pulse number of the
  128 pulses long sequence of pulses used to determine the spectral
  power). Transitions from the Q mode, during which there is a strong $P/P_3=0.08$ spectral feature, to the B mode, during which low-frequency modulation dominates, are evident.}
\end{figure}

As noted
before, the spectral features in the LRFS have a high quality factor,
suggesting a relatively stable periodicity. Nevertheless, we
observe significant broadening. The variability of $P_3$ can be
demonstrated more directly by exploiting the Sliding 2DFS (S2DFS;
\citealt{ssw09}), which has temporal resolution. The S2DFS is obtained
by computing the 2DFS for separate blocks of data (here component II,
for which both the modulation and intensity are strongest), where the
block-size is $n_\mathrm{dft}$ pulses of data. The first block
analysed contains the first $n_\mathrm{dft}$ pulses, and successive blocks
start one pulse later. Therefore, the results for each block are not
independent of each other. The spectra obtained for each block were
integrated over $P_2$, resulting in Fig. \ref{FigSlidingP3}. See
\cite{ssw09} for a full description of the method.

Figure \ref{FigSlidingP3} clearly shows mode changes between the
Q mode (strong spectral power at $P/P_3=0.08$ cpp) and B mode (strong
spectral power at lower frequencies). However, since the resolution is
limited by $n_\mathrm{dft}$ pulses, this method cannot be used to separate
the modes more accurately than the method described in
Sect. \ref{SectModeSep}. Despite the somewhat limited fluctuation
frequency resolution defined by $n_\mathrm{dft}$, the fluctuation
frequency is observed to be variable (see for example the increase in
fluctuation frequency at block number 4300).

The magnitude of the variability was quantified for the Q mode by
fitting the curves shown in Fig. \ref{FigSpectra} in the left
side-panels of the 2DFS of both profile components. The spectral
feature was fit with a Gaussian function, while the background was
described by a second-order polynomial. The $P/P_3=0$ bin and all bins
above and including the second harmonic were excluded. Although not
perfect, a Gaussian fit was deemed accurate enough to define the
width. We found that the width of the spectral feature, hence the
magnitude of the observed variability in $P_3$, is within the errors
identical in both components (the ratio is $1.07\pm0.06$). This is expected since phase-locking implies that the magnitude of the variations in $P_3$ is the same in the two components.

\subsection{$P_3$-fold}

So far, the subpulse modulation of PSR B1839--04 has been detected and
quantified using Fourier based techniques. A completely independent
method to visualise the time-averaged properties of subpulse
modulation is by folding the data at the period $P_3$. $P_3$-folding
essentially averages the pulse stack over the modulation cycle,
resulting in the average drift band shape without assuming a
sinusoidal modulation pattern. A complication is that $P_3$ is
variable throughout the observation, hence the fluctuations in $P_3$
have to be detected and compensated for. This allows longer stretches
of data to be folded, without the modulation cycle being smeared out
because of misaligned modulation cycles during the averaging
process. Since the spectral feature in the LRFS is indeed smeared out
over a finite range of $P_3$ (see Figs. \ref{FigSpectra} and \ref{FigSlidingP3}), this
correction can be expected to be essential to obtain a satisfactory
result for PSR B1839--04. $P_3$-folding has been done
before, but without resolving the $P_3$ variations
(e.g. \citealt{dr01,esv03}). The same algorithm as described below was
used by \cite{hsw+13}.

The process of $P_3$-folding starts with specifying the $P_3$ value used for the folding, here taken to be $12.4P$ as measured in
Sect. \ref{SectFluctModeSep}. Here, the 2005 data in the Q mode were
analysed. In PSRSALSA, the basic approach to fold the data is to
separate out blocks of continuous pulses (in this case taken to have a
length of 37 pulses, or about three modulation cycles). Then each block
is folded with the specified $P_3$ value (fixed-period folding). The
variability in $P_3$ is taken into account by finding appropriate
phase offsets as explained below. Analogous to what is done for the
fluctuation analysis, only stretches of continuous data with a
duration of a multiple of 37 pulses were used to ensure that the
pulses in each block of data being analysed is continuous (resulting
in 7141 usable pulses). Increasing the block-size would increase the
accuracy to correct for $P_3$ variability because of the increased
overall $S/N$ per block. However, an increased block size will result
in a pattern that is more smeared out because the fixed-period folding is not accurate enough within a block of data, hence a
compromise needs to be made.

\begin{figure}
\begin{center}
\includegraphics[height=0.99\hsize,angle=270]{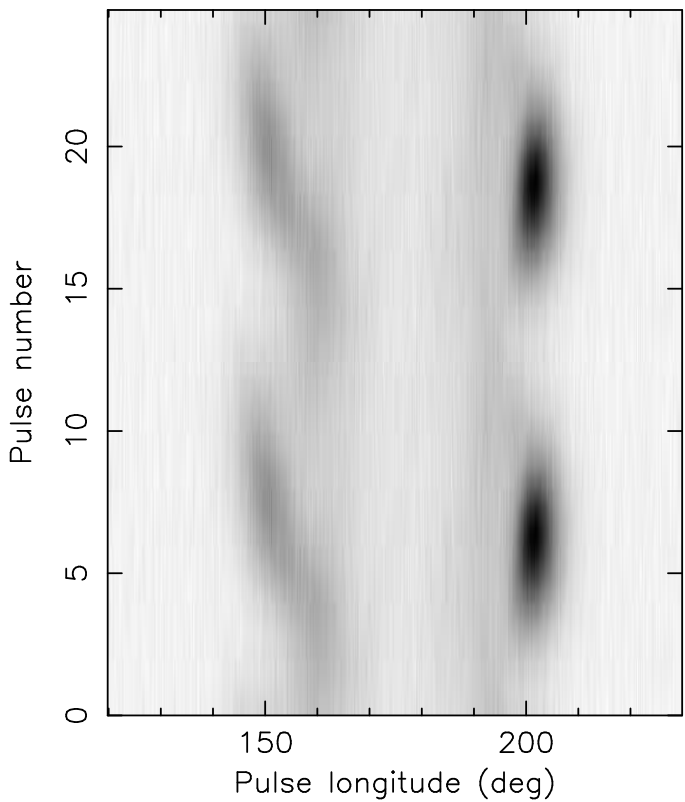}
\end{center}
\caption{\label{FigP3fold}$P_3$-fold of the added 2005 data of PSR
  B1839--04 in the Q mode. The average $P_3=12.4 P$ modulation cycle
  is shown twice for continuity.}
\end{figure}

The variations in $P_3$ are resolved by an iterative process during
which the modulation patterns obtained for the individual blocks of
data are compared, resulting in their relative offsets. This
means that if $P_3$ is
slightly larger than the typical value specified, the drift band will
appear slightly later in the second block than in the first. Hence
when folding the data, this additional offset has to be subtracted
first. These offsets are determined by cross-correlating the modulation pattern of the blocks of data and a template.
Only the on-pulse region, containing signal, was used to
correlate to determine the offsets, but the full pulse longitude
range is folded. Since the cross-correlation relies on knowledge
about the shape of the modulation pattern, an iterative process is
required. During the first iteration this knowledge is not available,
hence the modulation pattern of the first block of data is used to
determine the offset of the second block of data. After coherent
addition, this result is used to determine the offset of the
third block, etc. After folding the full data set using these offsets,
the result is the ``average drift band''. However, the determined
offsets can be refined by using this high $S/N$ average drift band as
a template to compute new cross-correlations, allowing more precise
offsets to be determined for each block. In total, four iterations
were used to calculate Fig. \ref{FigP3fold}. The result is shown twice
on top of each other for continuity.

The intrinsic resolution of the $P_3$-fold is one pulse period, but
``oversampling'' could give a cosmetically improved display of the result. This means
specifying more than $P_3/P$ bins covering the modulation cycle,
resulting in bins that are not completely independent of each
other. Here the $P_3$ cycle was divided into 50 equally spaced bins. A
given pulse was added to each of the 50 bins during the averaging
process after applying a weight from a Gaussian distribution with a
standard deviation of 4 bins and an offset corresponding to the
difference between the expected modulation phase of the given pulse
and that of the bin.  This Gaussian smoothing effectively makes the
obtained resolution of the modulation cycle equal to one pulse period.

Figure \ref{FigP3fold} shows the resulting $P_3$-fold in which the power
is concentrated in two pulse longitude ranges corresponding to components
Ia and IIb. The leading component clearly shows negative drift,
that is, subpulses moving towards the leading edge. The power in the
trailing peak does not clearly show drifting, consistent with the fact
that the drift-rate should be much lower. Nevertheless, the figure
reveals that there is drift such that at the trailing side the pattern
is brighter at later times (higher up along the vertical axis). This
is confirmed by analysing the centroid of the power in each column as
a function of pulse longitude. This confirms that indeed subpulses drift
in opposite directions in components Ia and IIb using a completely
different method independent of Fourier techniques.

\section{Polarization}
\label{SectPolarization}

Radio polarization, in particular the observed position angle (PA) of
the linear polarization, can be used to constrain the magnetic
inclination angle $\alpha$ of the magnetic axis and the inclination
angle $\zeta$ of our line of sight with respect to the rotation
axis. Here we apply the methodology described in \cite{rwj15} and
implemented in PSRSALSA on the 1408 MHz polarimetric pulse profile of
PSR B1839--04 published by \cite{gl98}, which is made available through
the European Pulsar Network Data
Archive\footnote{http://www.epta.eu.org/epndb/}, to derive
robust constraints after systematic consideration of the relevant
uncertainties.

\begin{figure}
\begin{center}
\includegraphics[height=\hsize,angle=270,trim=0 -3 0 -18.5,clip=true]{paswing.ps}
\includegraphics[height=\hsize,angle=270]{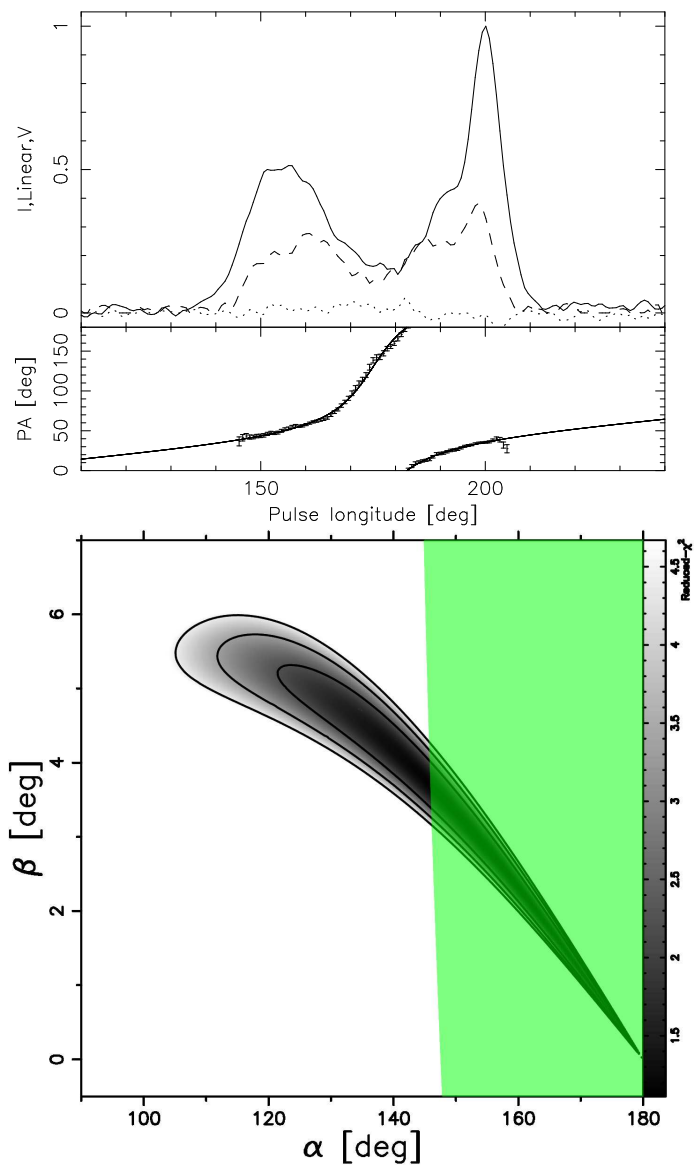}
\end{center}
\caption{\label{FigPolPlot}{\em Top panel:} Polarimetric pulse profile
  of PSR B1839--04 observed at 1408 MHz \citep{gl98} showing total
  intensity (solid line), linear polarization (dashed line) and
  circular polarization (dotted line). Below the profile the PA
  measurements are shown with the solid curve showing the best fit of
  the RVM. {\em Lower panel:} The goodness-of-fit of the RVM is shown
  in grey-scale, with the contours corresponding to 1, 2, and $3\sigma$
  levels. For the green area of the graph the observed pulse
  width is consistent with considerations of the expected beam size
  for this pulsar (see main text).}
\end{figure}

The top panel of Fig. \ref{FigPolPlot} shows the polarimetric profile
as derived by PSRSALSA from the recorded Stokes parameters. By default
the amount of linear polarization is computed using the method
described in \cite{wk74}. The PA points are only shown and used when
the linear polarization exceeds three times the standard deviation as
observed in the off-pulse region.

The pulse longitude dependence of the observed PA (i.e. PA-swing) can
be related to $\alpha$ and $\zeta$ through the rotating vector model (RVM;
\citealt{rc69a,kom70}). The observed PA-swing can be explained well
with this model, as shown in the top panel of Fig. \ref{FigPolPlot}
and is quantified by the reduced $\chi^2$ of 1.2. The geometrical
parameters of the RVM are not well constrained by the PA-swing, which
is, as is often the case, limited by the duty cycle of the profile. The
degeneracy in the model parameters is quantified in the lower panel of
Fig. \ref{FigPolPlot}, which shows the goodness-of-fit of the RVM as
function of $\alpha$ and $\beta$, where $\beta=\zeta-\alpha$ is the
impact parameter of the line of sight with respect to the magnetic
axis. The black contours correspond to $\chi^2$ levels that are two, three,
and four times higher than the lowest reduced-$\chi^2$, equivalent to the
1, 2 and $3\sigma$ levels\footnote{There are unmodelled features in the
  observed PA-swing when the reduced-$\chi^2$ exceeds 1, indicating
  that the derived uncertainties of the PA-values do not reflect the
  discrepancy between the data and the model. The fact that the model
  is inaccurate should be reflected in a larger allowed range in model
  parameters. This is achieved, at least to some level, by rescaling the measurement
  uncertainties to effectively include the
  estimated model uncertainty. 
Since the deviations are not uncorrelated Gaussian noise, a more realistic description of the model is required to quantify the uncertainties better.}. It can be seen that $\beta$ is
constrained to be positive, corresponding to a positive gradient of
the PA-swing. The fact that $\alpha$ is constrained to be larger than
$90\degr$ means for a positive $\beta$ that there is an ``inner'' line
of sight, or in other words, that the line of sight passes in between the rotational and
magnetic axis.

\begin{table}[!tb]
\caption{\label{TableWidthParameters}Measured parameters used to obtain the green region as indicated in Fig. \ref{FigPolPlot}. }
\begin{center}
\begin{tabular}{cccc}
\hline
\hline
$\phi_\mathrm{left}$ & $\phi_\mathrm{right}$ & $\phi_\mathrm{fid}$ & $\phi_\mathrm{0}$\\

[deg]   & [deg]& [deg]& [deg]\\
\hline
$141.5\pm2.9$ & $208.5\pm1.6$ & $175\pm25$ & $174.9^{+0.5}_{-0.4}$\\
\hline
\end{tabular}
\tablefoot{The parameters are the pulse longitudes of the left and right edge of the pulse profile, the position of the fiducial plane, and the inflection point of the PA-swing.}
\end{center}
\end{table}

\cite{rwj15} provided a more elaborate discussion about the used RVM fit
methodology and also discussed how beam opening
angle considerations in combination with constraints on aberration and
retardation effects and the observed pulse width can be used to obtain
a further constraint on $\alpha$ and $\beta$. The relevant input
parameters and their uncertainties can be found in Table
\ref{TableWidthParameters}. Here the pulse longitude of the left and
right edge of the pulse profile are defined at the 10\% peak intensity
level and the error bars are determined from the difference with those
at the 20\% level. The fiducial plane position, the pulse longitude
corresponding to the line of sight being closest to the magnetic axis,
was set to the mid-point between the left and right edge of the
profile. Since the profile is not perfectly symmetric, this is a
somewhat subjective choice. To account for a possible partially active
open field line region, the uncertainty was taken to be relatively
large such that the range of fiducial plane positions includes the
position of the two main profile peaks. This would allow for a
``missing'' component either before or after the observed
profile. Following the procedure outlined in \cite{rwj15}, only a
subsection of parameter space as indicated by the green region in
Fig. \ref{FigPolPlot} is consistent with the predicted open field line
region of a dipole field.

Figure \ref{FigPolPlot} suggests that $\alpha$ is likely to be
relatively small (i.e. close to $180\degr$). The magnetic inclination
angle is likely to be smaller than $35\degr$. This is because the
observed pulse width is relatively wide ($w_{10}=67\degr$), while
there is no evidence for a large emission height from the offset
between the fiducial plane position and the inflection point of the
PA-swing \citep{bcw91}. In fact, the measured value of the inflection
point slightly precedes the assumed fiducial plane position,
although not significantly so. Since the opposite is expected, it
suggests that if the fiducial plane position is close to the defined
central point in the pulse profile, the emission height should be
low. In that case the pulsar will be a relatively aligned rotator with
$\alpha$ close to $180\degr$. A small inclination angle is consistent
with the analysis of \cite{whx02}, for instance,
who found
$\alpha=7.4\degr$ without a defined uncertainty.

\section{Discussion}
\label{SectDiscussion}

In this section the bi-drifting phenomenon is discussed by first of
all comparing the two known bi-drifters. Explaining these results is
challenging, and a range of ideas, some specifically proposed in the
literature as a model for bi-drifting, are argued to be
implausible. Finally, potential ways to avoid these theoretical
problems are identified.

\subsection{Comparison with PSR J0815+0939}

As noted above, bi-drifting was reported for one more pulsar: J0815+0939
\citep{mlc+04,clm+05}.  With a shorter period ($P=0.645$ seconds), but
an almost four times higher spin-down rate, the characteristic age of
PSR J0815+0939 ($\tau_c=74$ Myr) is similar to that of B1839--04
($\tau_c=57$ Myr). This places both pulsars at the lower spin-down
rate end of the distribution of known pulsars in the $P-\dot{P}$
diagram. Nevertheless, their spin-down properties are by no means
exceptional, making it unlikely that spin-down parameters alone
determine if pulsars show bi-drifting.

In addition to the bi-drifting, another interesting feature about PSR
J0815+0939 is its distinct pulse profile shape at a frequency of 327
and 430 MHz, which has four very clearly separated components with
deep minima between them \citep{clm+05}. However, at 1400 MHz the
profile is comparable to that of PSR B1839--04 at a similar frequency,
with overlapping profile components resulting in a much more double-peaked structure. At 408 MHz the profile of PSR B1839--04 shows no
evidence for a similar frequency evolution as J0815+0939
\citep{gl98}, although it cannot be ruled out that at even lower
frequencies something similar might occur. Nevertheless, our analysis
does show the presence of four profile components in the profile of
PSR B1839--04.

The profile of PSR J0815+0939 is more narrow when expressed in
seconds. \cite{clm+05}, for example, reported ${w_{10}=183}$ ms at 430 MHz. However, it is more
relevant that its duty cycle is very large, almost 28\% and slightly more
at 1400 MHz. This is even larger than the duty cycle measured here for PSR~B1839--04 at 1380 MHz (19\%).
This suggest that PSR~J0815+0939 is a relatively aligned rotator,
similar to PSR~B1839--04. A difference is that the profile of PSR~J0815+0939 appears to narrow at lower frequencies, while that of PSR~B1839--04 widens (e.g. \citealt{cw14}).

Subpulse modulation is observed for all four components of PSR
J0815+0939. The first component has an unclear drift sense (appears to
change during observations). The second component has positive drift,
while the remaining components have negative drift. This feature
is reported to be stable from observation to observation
\citep{clm+05}. It must be noted that a more detailed analysis of the
first component of PSR J0815+0939 might reveal if there is any
systematic drift direction detectable. Therefore it is unclear if the
first component of PSR J0815+0939 is comparable with the second
component of PSR B1839--04 in the sense that at first glance the drift
sense is unclear. The bi-drifting in both pulsars is consistent with a
symmetry such that the drift sense is opposite in the two halves of
the profile, but the same within a given half of the profile. This
could well be a generic feature of the bi-drifting
phenomenon. 

\subsection{Basic requirements for a model for bi-drifting}

Any model for bi-drifting should be able to explain at least the
following basic observed properties.
\begin{enumerate}
\item Why different profile components can have subpulses drifting in opposite directions.
\item Why this is so uncommon in the overall pulsar population.
\item Why $P_3$ is the same in the different components.
\item How very strict phase-locking of the modulation cycle in different components can be maintained.
\end{enumerate}
Especially the last point is a critically difficult property to
be explained by emission models. Phase-locking is explicitly demonstrated
here for PSR B1839--04 over a timescale of years. In addition, it is
unaffected by disruptions by mode changes.

\subsection{Theoretical difficulties in interpreting bi-drifting}

As mentioned in the introduction, a widely used model to explain
drifting subpulses is the carousel model. This is a specific physical
model based on circulation of pair-production sites (known as sparks),
responsible for beamlets of radio emission. The circulation around the
magnetic axis is caused by an $\bm{E}\times\bm{B}$ drift. Although
much of the following discussion is described in terms of this model,
the arguments in this subsection apply equally to any model that tries
to explains drifting subpulses by a circulation of radio beamlets
around the magnetic pole.

\subsubsection{Oppositely circulating nested beamlet systems}
\label{SectOppositelyCirculatingBeamlets}

\cite{qlz+04} argued, motivated by the results for PSR J0815+0939, that
bi-drifting can be explained in the context of the inner annular gap
(IAG) discussed by \cite{qlw+04}. Within the framework of the
\cite{rs75} model, particle acceleration is expected in the charge-depleted region in the open field line zone close to the neutron star
surface (polar gap). This polar gap is divided into two regions: a
central region (the inner core gap; ICG) above which the charge
density of the magnetosphere has an opposite sign compared to that
above the annular region surrounding it (IAG). The IAG is likely to
require the star to be a bare strange star in order for it to
develop. It was argued by \cite{qlz+04} that a carousel of sparks could
exist in both regions, each rotating in an opposite sense. This
provides the physical basis for different components having opposite
drift senses.

As pointed out by \cite{vt12}, the \cite{rs75} model without the
addition of an IAG does not necessarily imply that only one sense of
carousel rotation is allowed. Therefore without invoking emission from
the IAG, two nested beamlet systems rotating in opposite directions
could be considered. The arguments made below about the \cite{qlz+04}
model therefore equally apply to such a model, or any model based on
circulation occurring in opposite directions.

In its most simple form, this model would predict that there is a pair
of inner components with the same drift sense, and an outer pair both
having the same drift sense (opposite from the inner pair). This is
inconsistent with the patterns observed for both bi-drifting
pulsars. Therefore a symmetry-breaking mechanism needs to be invoked to
change the natural symmetry implied by the IAG model. \cite{qlz+04}
proposed that emission produced at multiple emission heights in
combination with delays arising from retardation and aberration could
be responsible. For PSR J0815+0939 it was proposed that for both the
IAG and ICG radio emission is generated at two separate heights,
giving effectively rise to four systems of rotating radio beamlets,
the higher emission height producing a copy of the lower emission
height system of rotating beamlets with a larger opening angle. For a
relatively central cut of the line through the emission beam, eight
profile components are expected. It was argued that for each beamlet
system only the trailing component is bright enough to be observed and
that the emission height difference of the IAG compared to the ICG is
large enough to explain the observed pattern of drift senses for PSR
J0815+0939. Especially now that two pulsars show a similar ``unnatural''
symmetry in the observed drift senses of the different components,
such a scenario appears to be unlikely.

An even more fundamental problem of such a model is that $P_3$ is not
required to be the same for components showing opposite drift
senses. This is first of all because the two beamlet systems not
necessarily have the same number of beamlets. Secondly, the
circulation time of the two beamlet systems cannot be expected to be
the same. In the model of \cite{qlz+04}, $\bm{E}\times\bm{B}$ can be
expected to be different for the two carousels (see also their Eq. 2)
because $\bm{B}$ will be slightly different in the two gaps, but in
addition, $\bm{E}$ will not be the same\footnote{Since for a start the
  sign of $\bm{E}$ needs to be opposite to explain the opposite drift
  sense, there is no a priori reason why the electric fields in the
  two gaps should have exactly the same magnitude.}.  \cite{qlz+04}
simulated the drifting subpulses of PSR J0815+0939 with two carousels
containing 21 and 32 sparks. This implies that the circulation time
needs to be different, and fine tuned, to make the observed $P_3$ to
be the same for the different components.

It is very hard to imagine a way to keep the extremely strict phase
relation between the two carousels that are circulating with
different speeds in opposite directions. As argued by \cite{qlz+04},
non-linear interactions between the two gaps might potentially lead to
frequency-locking into a rational ratio. Even if such a mechanism
exists, it would be a coincidence that the apparent $P_3$ is the same
for both carousels. The fact that both known bi-drifting pulsars show
that $P_3$ is identical and now phase-locking is explicitly
demonstrated for PSR B1839--04, it is very unlikely that oppositely
rotating carousels are responsible. It must be stressed that any model
with oppositely circulating beamlet systems faces the same problems,
hence other routes need to be explored to explain bi-drifting.

\subsubsection{Localised beamlet drift}

In the model of \cite{jon13}, subbeam circulation is not caused by an
$\bm{E}\times\bm{B}$ drift. Therefore, if a circulating beamlet system
were to exist, it could rotate in both directions. This is
in contrast to the \cite{rs75} model, which in its basic form does not allow
circulation in both directions (although again, see
\citealt{vt12}). Moreover, as pointed out by \cite{jon14}, his model
does not require that subbeams are organised in a rigid circular
pattern. The drift of subbeams can be a localised effect with sparks
moving along possible open paths in any desired direction. Within this
framework, bi-drifting is not problematic, and in \cite{jon14} a
possible subbeam configuration for PSR J0815+0939 is presented.

As for the scenario discussed in
Sect. \ref{SectOppositelyCirculatingBeamlets}, the increase in
flexibility in the model leads to fundamental problems in explaining
the observational results. Why is bi-drifting so uncommon? Why is
$P_3$ the same in the different components and how can the modulation
be phase-locked over years? Any model that attempts to explain the
oppositely drifting subpulses by two separate localised systems
requires a mechanism to synchronise the modulation, thereby making it
effectively a single system.

\subsubsection{Alias effect of circulating beamlets}

A fundamental problem so far in interpreting bi-drifting has been the
observed phase-locking. For pulsars without bi-drifting
phase-locking of modulation observed in different components resulting
in identical $P_3$ values is very common
(e.g. \citealt{wes06,wse07}). When multiple carousels are required to
reproduce the data, phase locking can most easily be achieved by
having two subbeam systems rotating with the same period and the same
number of sparks in effectively a single locked system. The two nested
rings of emission might be out of phase such that a spark on the inner
ring is kept in place through an interaction with the preceding and
following sparks on the outer ring by maintaining a fixed separation.

Such a phase-locked system with two carousels circulating in the same
direction has been successfully used to model the drifting subpulses
of for instance PSR B0818--41 \citep{bggs07,bgg09} and PSR B0826--34
\citep{ggks04,elg+05,bgg08}. Moreover, both these pulsars show reversals of
the observed drift direction such that the observed drift sense is
different at different times. Some of these authors attribute the apparent
drift reversals to an alias effect caused by a change in circulation
period, without an actual reversal of the circulation. Alias can
occur when a periodicity (here the circulation) is sampled with a
different periodicity (here the stellar rotation). To give a numerical
example: if each beamlet rotates by 0.1 beamlet separation per stellar
rotation, the observed $P_3$ would be $10P$. However, if in the same
amount of time each beamlet rotated by 0.9 beamlet separation,
the observed $P_3$ would also be $10P$, but the observed drift sense
would be opposite. The two cases correspond to two different alias
orders. This means that an observed reversal of drift direction does
not imply that the subbeam system reverses its sense of circulation.

This raises the question as to whether bi-drifting can be explained by having the
leading and trailing half of a carousel be in a different alias
order, or whether two phase-locked carousels can have a different alias
order. The beamlets in a single circulation system could in principle
move with different angular velocities in different parts of the
carousel. However, to conserve the number of beamlets (otherwise
phase-locking cannot be maintained), any increase in average angular
velocity at a given location should be balanced by an increase in
beamlet separation. As a result, both $P_3$ and the alias order will be
identical. This is a powerful prediction of a circulation model that
makes it a good model to explain the phase-locking as observed for most
pulsars, but it makes such a model unsuitable to explain bi-drifting.

A scenario with two carousels having different alias orders is
also highly unlikely. First of all, as for the scenario discussed in
Sect. \ref{SectOppositelyCirculatingBeamlets}, the wrong type of
symmetry is expected with each pair of components corresponding to a
single carousel having the same drift sense. Secondly, in
Sect. \ref{SectPhaseLocking} we demonstrated that the amount of
variability in $P_3$ is, within the errors, identical in the two
components. In a circulation scenario this suggests that the
underlying circulation period is the same as well,
since increasing the alias order would result in more $P_3$ variability for a given fractional change of the circulation period.
This makes it unlikely that the alias order is different
within a phase-locked system.

There is an even more fundamental problem with such a double carousel
interpretation. To have two phase-locked carousels implies that the
rotation period of both carousels should have a fixed ratio (with
a ratio of one being the most natural choice to explain phase
locking). In principle,  another free parameter is the number of
beamlets on each carousel. To have different alias orders for the two
carousels implies that the carousel rotation period and the number of
beamlets cannot be the same for both carousels. Therefore, fine-tuning
is required to make $P_3$ the same for all the components as is
observed for both bi-drifters. This  makes this scenario highly improbable.

\subsubsection{Alias effect in a pulsation model}

The theoretical ideas discussed so far assume that a circulation of
subbeams is responsible for subpulse drift. However, this is not the
only geometrical configuration that could be the
explanation. \cite{cr04a} demonstrated that drifting subpulses can be
explained with non-radial pulsations, an idea already considered by
\cite{dc68}. In such a model $P_3$ is determined by the beat frequency
between the oscillation period and the stellar rotation period. The
observed phase-locking and the fact that $P_3$ is identical strongly
suggest that in such a model the oscillation frequency responsible
for the drifting subpulses should be the same in each profile
component. This implies that the alias order should also
be the same, making
it impossible to explain bi-drifting in this scenario.

\subsubsection{Multiple projections of a single carousel}

It is, at least in principle, possible that a single carousel produces
two beamlet systems. For instance this could occur if radio emission
is produced at two separate heights on the same field lines (see also Sect. \ref{SectOppositelyCirculatingBeamlets}). This
would result in two sets of circulating beamlets that are
phase-locked, but with a different opening angle with respect to the
magnetic axis. These two radio projections of the same carousel are
magnified by different amounts because of the curved magnetic field lines. The two projections will very naturally
result in identical observed $P_3$ values and phase-locking. However,
the drift sense should be the same in all components. 

An alternative physical mechanism for producing multiple projections would
be refraction in the magnetosphere. When refraction occurs towards the
magnetic axis by a large enough amount, beamlets produced by a spark
could be visible at the other side of the magnetic pole (see
e.g. \citealt{pet00,wsv+03}). If the refraction confines the radiation
to the planes containing the dipolar field lines, the resulting
beamlets appear $180\degr$ out of phase in magnetic azimuth (see also
\citealt{esv03}). However, in this scenario the drift sense is
not affected either, hence it cannot be used as a model for bi-drifting.

\subsubsection{Multiple projections by two magnetic poles}

Some pulsars show both a main- and interpulse, which in general are
thought to be produced at opposite magnetic poles. There are two
pulsars known with an interpulse showing the same $P_3$ modulation and
phase-locking compared with the main-pulse: PSRs B1702--19
\citep{wws07} and B1055--52 \citep{wwj12}. Inspired by these results,
a different way of generating multiple projections can be considered with opposite
magnetic poles producing a carousel with the same number of sparks and
the same rotation period. In addition, the phase-locking suggests
there must be communication between the two poles, making it
effectively a single system. How this communication can be established is by no means obvious
(see the two mentioned papers and references therein).

Could bi-drifting be explained by one projection being formed from the
carousel at the far side of the neutron star? This is in principle
possible if radio emission is also generated directed towards
the neutron star (e.g. \citealt{dzg05}). If the magnetic inclination
angle is not very close to orthogonality, as is likely in the case of both
pulsars showing bi-drifting, each magnetic pole is only observed once
per stellar rotation, one as outward-directed radiation and the other as inward-directed radiation. Both poles can be expected to be observed at roughly the same rotational phase. One difficulty with such a model is to explain
why bi-drifting is not observed more often since the same argument could apply to all pulsars. Secondly, even in this
scenario no bi-drifting is expected to be observed. This is
because in a circulation model based on \cite{rs75} the
$\bm{E}\times\bm{B}$ drift is expected to be such that the two
carousels rotate in the same direction so that the sparks at the two
poles can be imagined to be connected by rigid spokes through the
neutron star interior\footnote{This follows from the fact that the
  electric field structure is the same with respect to the neutron
  star surface, while $\bm{B}$ is pointing into the surface for one
  pole, but out of the surface for the other.}. In such a scenario,
assuming a roughly dipolar field, for any possible line of sight the
drift sense is expected to be the same for the visible parts of the projections produced by both
poles. For such a scenario to work, it requires additional effects
that were not considered, such as potentially gravitational bending
or distortions of the projection through magnetospheric plasma interactions (e.g. \citealt{dfs+05}).

\subsection{What could work?}

After pointing out how problematic explaining bi-drifting is with any
geometric model, even after ignoring the question whether such a
configuration is physically plausible, are there any
possibilities left?

If drifting subpulses are caused by circulation, the following
requirements appear to be unavoidable to ensure phase-locking with
identical $P_3$ values in different components:
\begin{enumerate}
\item There must be circulation along a closed loop(s).
\item If there are multiple loops involved, they must be phase-locked and therefore rotating in the same direction.
\item Continuity of subbeam ``flux'': on average the same number of subbeams per unit time should pass a given point along the circulation path.
\end{enumerate}
The last point allows $P_2$ to be different for different components
by having a different separation between the subbeams at different
points along their path, while maintaining the same observed $P_3$
value. This also implies that the alias order is the same along the
path.

Closed loops are not necessarily required to maintain phase-locking in
different components. It could be imagined that there is a single point
in the pulsar magnetosphere, for example the centre of the magnetic
pole, driving structures away from the magnetic axis simultaneously in
different directions. This could result in bi-drifting without the
need for closed loops. It is not clear which physical mechanism
could be responsible, and therefore this scenario is not discussed in more detail.

\begin{figure}
\begin{center}
\includegraphics[width=\hsize,angle=0]{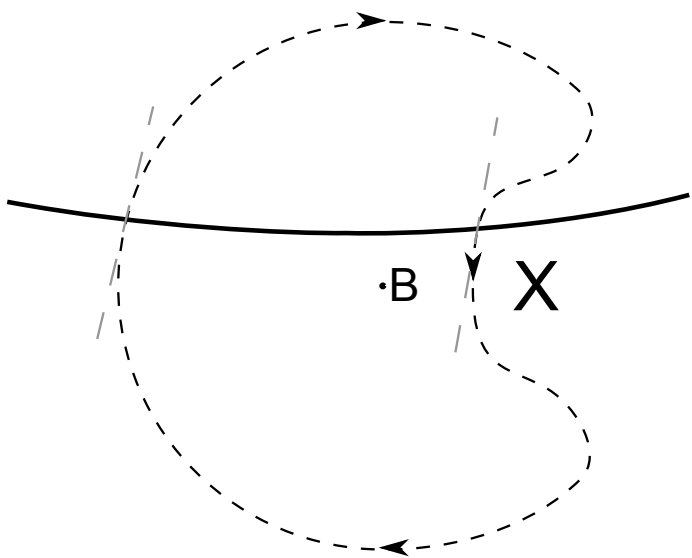}
\end{center}
\caption{\label{FigHorseShoe}Example of a possible closed subbeam
  circulation loop (dashed line) around the magnetic pole (labelled
  $\bm{B}$) that would result in bi-drifting as observed by an
  observer whose line of sight samples a path indicated by the thick
  solid line. The rotational axis is somewhere above the magnetic axis
  (not shown). To observe bi-drifting, the tangents of the path at
  the intersection points (grey dashed lines) should be roughly parallel, unlike what
  would occur for a circular path. The X indicates the position of a possible obstruction responsible for the distorted circulation path.}
\end{figure}

Subbeam circulation configurations exist that meet the above
requirements and could explain bi-drifting. An example of such a
configuration is shown in Fig. \ref{FigHorseShoe}. Drifting subpulses
can be expected to be observed at all intersection points of the line of sight with
the subbeam path where the two are not perpendicular to each other
(otherwise longitude stationary on/off modulation would be
observed without any drift). For a circular subbeam configuration the
tangents to the subbeam path will be mirror-symmetric with respect to the
meridian connecting the magnetic and rotation axis (vertical through the
middle of the figure). This would mean that the same drift direction
would be observed for two intersection points. Any path for which the two
tangents are roughly parallel rather than mirror-symmetric would result in bi-drifting.

Any closed loop with the same tangents at the intersection points with the line of sight would
result in identical observed drifting subpulse properties, hence
Fig. \ref{FigHorseShoe} should be considered as an example of a
possible configuration. The example in Fig. \ref{FigHorseShoe} might
result from a distortion in the polar cap shape, forcing the subbeams
to circulate around an obstruction (indicated with an X). For more realistic magnetic field
line configurations the polar cap is indeed non-circular and a notch \citep{dhr04} is expected. However, since
both bi-drifters are likely to have relatively small magnetic
inclinations, this effect cannot be expected to be particularly strong
for these pulsars. Particularly large departures of an
idealized  dipolar field structure might be responsible for the deformed
circulation path\footnote{Here it should be noted that the \cite{rs75}
  model relies on multipoles to be significant to increase the curvature of the field lines.}. The rarity of
bi-drifters might indicate that there is a fine balance between having
a large enough deformation without preventing regular circulation from
being maintained. Possibly the mode changes are a consequence of the
circulation failing at certain times. The observed drift rate as function of rotational phase, as can be quantified using the subpulse phase track, might in principle be used to quantify the shape of the circulation path and might be used to relate this to the required distortion in the magnetic field.

\section{Summary and conclusions}
\label{SectConclusions}

A new set of software tools, available as a new open-source
data-analysis project called PSRSALSA, were described and its power is
illustrated by analysing the radio data of \mbox{PSR B1839--04}. In
particular, the highly unusual phenomenon of bi-drifting was
quantified in detail. Bi-drifting is the phenomenon that the drift
direction of subpulses is different in different pulse profile
components, which is very hard to explain theoretically.

A robust constraint on the magnetic inclination angle $\alpha$ was
derived from archival radio polarization measurements after systematic
consideration of the relevant uncertainties following the methodology
described in \cite{rwj15}. PSR~\mbox{B1839--04} is likely to be relatively
aligned ($\alpha$ is constrained to be smaller than $35\degr$). It is
likely that the only other known bi-drifter, PSR J0815+0939, shares
this property. Another similarity is that the wide profiles of both
pulsars appear to be composed of four components.

WSRT observations revealed that two distinct emission modes operate in
this pulsar, with the bi-drifting periodic subpulse modulation being
present only during the Q mode. During the B mode, no regular drifting
subpulses are observed, the pulse profile is more intense, and its
components are closer to each other.

The bi-drifting in PSR B1839--04 is not clearly visible in a
pulse stack.  Nevertheless, utilising high-$S/N$ mode-separated data
for the Q mode and sensitive analysis techniques, the existence of
bi-drifting was convincingly demonstrated by exploiting the modulation power in the second
harmonic observed in fluctuation
analysis. Furthermore, bi-drifting was confirmed using
$P_3$-folding. This is a completely independent method to visualise
the subpulse modulation pattern by averaging over all pattern
repetition periods $P_3$.

Three additional properties related to bi-drifting need to be explained
by a viable model for subpulse modulation. First of all, while
drifting subpulses are very common in the pulsar population, there are
only two known examples of pulsars showing bi-drifting. This makes
bi-drifting extremely rare. Secondly, for both bi-drifting pulsars
$P_3$ is identical for all profile components. Thirdly, both
bi-drifting pulsars are consistent with a symmetry such that all
components in each half of the profile show the same drift sense, or
no clear drift.

The modulation cycle of PSR B1839--04 is somewhat unstable, with
significant variability in $P_3$ being detected. The variability in
$P_3$ was shown to be identical in both components. Further analysis
firmly established that the modulation pattern responsible for the
bi-drifting is strictly phase-locked over a timescale of years. This
means that if the modulation cycle is slow in one component, it is
equally slow in the other such that the modulation patterns of both
components stay in phase. The phase-locking is not disrupted by the
mode changes, even though the modulation disappears during the
B mode. This implies that a single physical origin is responsible for
the modulation pattern observed in both components.

Phase-locking is critically difficult to explain by
models. As a result, many ideas, some specifically proposed in the
literature as a model for bi-drifting, were argued to be
implausible. These include any model that requires oppositely
circulating nested beamlet systems and any model invoking alias
effects in the framework of nested beamlet systems or non-radial
pulsations. Any model based on separate localised systems to explain
the oppositely drifting subpulses requires communication between these
systems to maintain synchronicity, effectively making it a single
physical system. Phase-locking of subpulse modulation is reported for
opposite magnetic poles of some pulsars, and the radiation of both
poles could in principle be observed in a single on-pulse window if
radiation is allowed to be directed towards the neutron star. However,
such a scenario does not naturally lead to bi-drifting. The same is
true for multiple projections of a single beamlet system, even if for
instance magnetospheric refraction causes radiation to cross the
magnetic axis.

We argued that within the framework of circulating beamlets
scenarios exist that could explain drifting subpulses. They require
subbeams to circulate around closed loops above the magnetic pole. For
bi-drifting to occur, the loop must be distorted such that the tangents
to the circulation path are roughly parallel rather than
mirror-symmetric with respect to the meridian connecting the
magnetic and rotation axis. This might reflect distortions in the  magnetic field
structure. It is possibly difficult to distort the circulation path
enough for bi-drifting to occur without disrupting circulation altogether, which would explain the rarity of bi-drifting.

\begin{acknowledgements}
The author is grateful for useful discussions, suggestions for
improvements and bug reports by Ben Stappers, Geoff Wright, Maciej
Serylak, Neil Young and Simon Rookyard. The Westerbork Synthesis Radio
Telescope is operated by the ASTRON (Netherlands Foundation for
Research in Astronomy) with support from NWO.

\end{acknowledgements}

%
%

\bibliographystyle{aa}

\end{document}